 \documentclass{emulateapj}

\usepackage{graphicx}
\usepackage{epstopdf}
\usepackage{epsfig}
\shorttitle{DASCH $10-100$ yr variable K2 giants} \shortauthors{Tang
et al.}

\begin{document}


\title{DASCH Discovery of Large Amplitude $\sim$10-100 Year
Variability in K Giants} 


\author{Sumin Tang\altaffilmark{1}, Jonathan Grindlay\altaffilmark{1},
Edward Los\altaffilmark{1}, Silas Laycock\altaffilmark{2}}
\altaffiltext{1}{Harvard-Smithsonian Center for Astrophysics, 60
Garden St, Cambridge, MA 02138} \altaffiltext{2}{Gemini Observatory,
670 North Aohoku Place, Hilo, HI 96720}

\email{stang@cfa.harvard.edu}


\begin{abstract}
Here we present the discovery of three unusual long-term variables
found in the Digital Access to a Sky Century at Harvard (DASCH)
project, with $\sim$1 magnitude variations in their lightcurves on
$\sim10-100$ yr timescales. They are all spectroscopically
identified as K2III giant stars, probably in the thick disk. Their
lightcurves do not match any previously measured for known types of
variable stars, or any theoretical model reported for red giants,
and instead suggest a new dust formation mechanism or the first
direct observation of ``short" timescale evolution-driven
variability. More theoretical work on the lithium flash near the Red
Giant Branch (RGB) bump and the helium shell ignition in the lower
Asymptotic Giant Branch (AGB), as well as long term monitoring of
K2III thick disk stars is needed.

\end {abstract}


\section{Introduction}
The time domain, especially on 10-100 yr timescales, is poorly
explored despite its astrophysical importance. The Harvard College
Observatory (HCO) maintains a collection of more than 500,000 glass
astrophotographic plates from the 1880s to the 1980s, constituting
the only continuous record of the whole sky in existence. Every
point on the sky has been observed between 500 and 1000 times. This
100 years coverage is a unique resource for studying temporal
variations in the universe on $\sim$10-100 yr timescales. The
Digital Access to a Sky Century at Harvard (DASCH) collaboration has
developed an ultra-high speed digital plate scanner (Simcoe et al.
2006), and will ultimately enable the full Harvard plate collection
to be digitized. We have developed the astrometry and photometry
pipeline, and scanned 7000 plates in six different fields. An
overview of the DASCH project is presented in Paper I (Grindlay et
al. 2010; see also Grindlay et al. 2009), and the photometry and
astrometry pipelines are described in paper II (Laycock et al. 2010)
and paper III (Tang et al. 2010a).

Here we present the discovery of three unusual long-term variables
found in the DASCH scans near open cluster M44, which showed $\sim$1
mag dimmings and recoveries on $\sim10-100$ yr timescales in their
lightcurves. Such variations are very unusual and haven't been seen
in any other common classes. We present their lightcurves and
spectra in section 2. Discussion on individual objects are in
Section 3 and summary is in Section 4.

\section{Discovery of Three Unusual Variables}

\subsection{Candidate Selection}
Three unusual long-term variables presented here were found from
$\sim400$ variables found on $\sim 1200$ M44 plates by their
peculiar long-term variabilities. These plates cover 5$-$25 degrees
on a side with typical limiting magnitudes $14-15$ mag (Laycock et
al. 2010). There are $\sim1.2\times10^5$ objects with more than 100
magnitude measurements. Details of our variable selection procedure
and general properties of variables found in DASCH scans near M44
are described in Tang et al. (2010b).

\subsection{DASCH lightcurves}
These variables showed unusual $\sim$1 mag dimmings on timescales
from 10 to 100 yr in their lightcurves, as shown in black dots in
Figure 1. DASCH J083038.5+140713 (hereafter J0830; named by its
equatorial coordinate in J2000; GSC2.3.2 catalog name N2313102243)
declined for 1 mag in a century. It is classified as a `MISC'
variable in ASAS (ASAS J083038+1407.3; Pojmanski, G. 2002) since it
became 0.3 mag brighter in V gradually from 2003 to 2007, and then
became 0.1 mag fainter from 2008 to 2009. DASCH J075445.9+164141
(hereafter J0754; GSC2.3.2 name N2211330177; ASAS J075446+1641.7)
showed a sharp decrease around 1930, and then slowly recovered in 10
years. Another dip was shown around 1892, but unfortunately we can
not constrain the lightcurve profile of the dip due to the lack of
data. DASCH J073606.5+211411 (hereafter J0736; GSC2.3.2 name
N2230030699; ASAS J073607+2114.2) showed a 1 mag dip from 1930s to
1950s. Both J0754 and J0736 are new variables found with DASCH.

\begin{figure*}
\epsfig{file=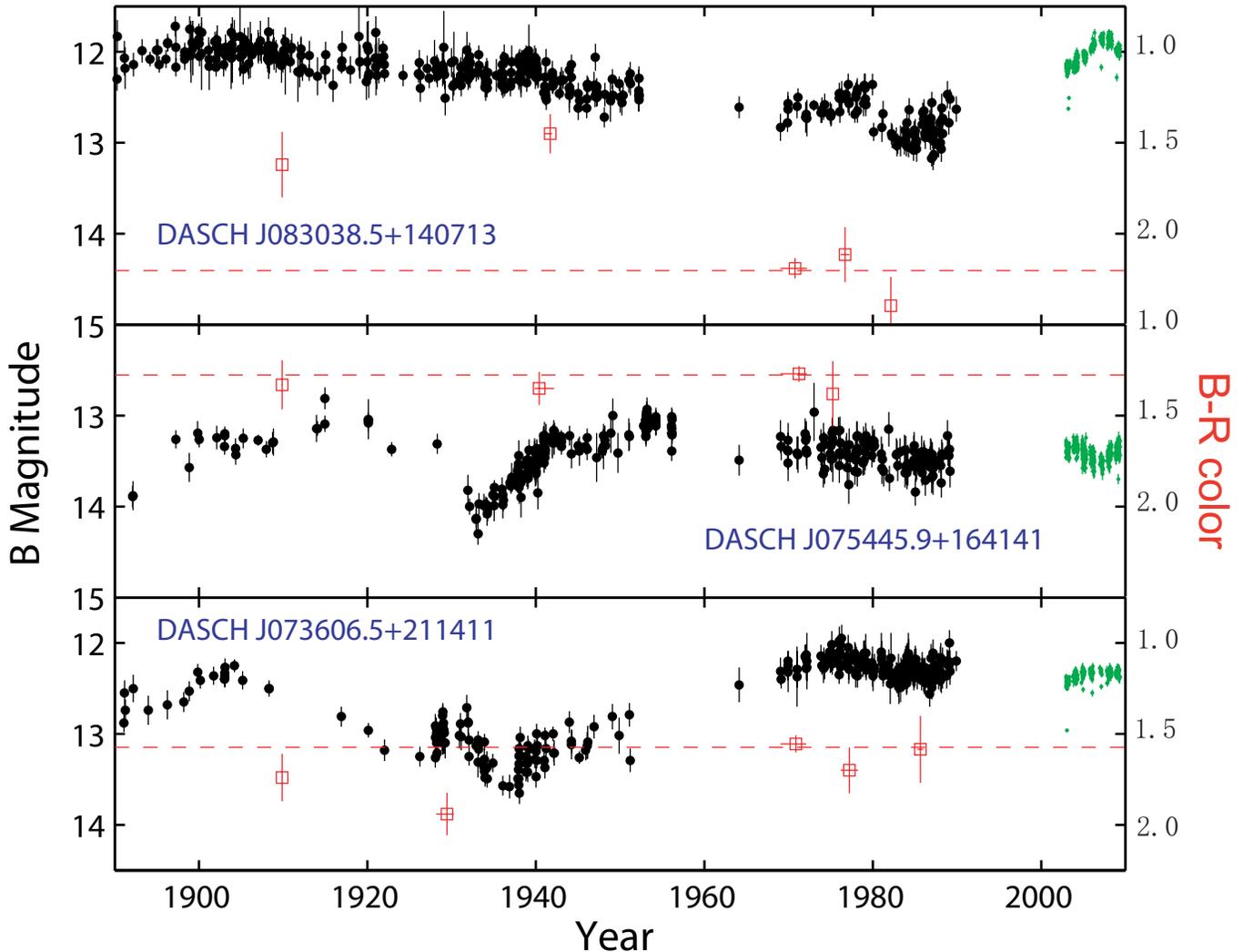, angle=0, width=\linewidth}
\caption{Lightcurves and color evolution of 3 unusual long-term
variables which were found in DASCH scans near M44. Black dots with
errorbars are the lightcurves from DASCH, small green dots are the
lightcurves from ASAS. Since ASAS data are in V band, while DASCH
magnitudes are B, we added 1.16 mag to the ASAS V magnitudes in the
plots which is the mean B-V value for K2III stars (Cox 2000). Red
open squares are the B-R color derived from plates with y-axis
labeled in the right, and red dashed lines mark the weighted mean
B-R color values from 1970s to 1980s.}
\end{figure*}

\subsection{Possible Color Evolution Derived from Plates}
The lightcurves of color variations back in time would constrain
variable extinction, but is difficult to derive from DASCH. The
majority of the Harvard plate collection are blue sensitive plates,
and a small fraction of plates used filters to produce red and
yellow sensitive measurements with details of wavelength responses
unavailable. In order to generate consistent magnitudes, we did
color-term fitting for the plates in annular bins to derive the
effective color-term $C$ in plates (Laycock et al. 2010), where $C$
of a given plate is defined by
$$m = B + C(B-R),$$
where $m$ is the effective magnitude in the plate, $B$ is the GSC2
$B$ magnitude and $R$ is the GSC2 $R$ magnitude. We then derive the
(B-R) color of a given object by comparing its magnitudes with its
neighbor stars (with B-R color from GSC2 catalog) in pairs of blue
vs red/yellow plates taken from the same night or very close in time
(mostly within a week). Each pair of plates gives a (B-R) color of
the object at that time. More details are described in Tang et al.
(2010a).

We bin some plate pairs close in time, and plot the color evolution
of the three variables by the red open squares in Figure 1. The
color evolution data are limited mainly by the small number of red
and yellow plates usually available. Both J0830 and J0736 are redder
when they are fainter, at 7$\sigma$ and 3$\sigma$ level,
respectively. We didn't detect any color change in J0754, although
we do not have red or yellow plates available during its major
dimming phase (1930-1940).

\subsection{Spectroscopic Observations}
Spectra were acquired with FAST spectrograph on the 1.5 m
Tillinghast reflector telescope at the F. L. Whipple Observatory
(FLWO) and GMOS long-slit spectroscopy blue channel on Gemini North.
They are wavelength-calibrated with standard packages, and are shown
in Figures 2 and 3. According to their spectra, they are all K2-type
stars. By comparing the region between $4900-5200$ \AA \ with FAST
spectra of several K2 standards with different luminosity classes,
we found all of them are giants (luminosity class III; Estimated
uncentainties in luminosity class are II-IV. Luminosity class I and
V are ruled out). We estimated metallicities [Fe/H]$\sim -0.3\pm0.3$
for J0830, and [Fe/H]$\sim -0.9\pm0.4$ for J0754 and J0736, by
comparing with FAST spectra of several standard K giant stars with
known metalicities (Faber et al. 1985). We also found that J0830 and
J0754 do not show velocity changes within measurement errors
($\sim8$ km/s), while J0736 showed significant radial velocity
changes in three different epochs, i.e. $11\pm6$ km/s on Feb 4th,
$-18\pm6$ km/s on Feb 19th, and $22\pm6$ km/s on April 18th, 2009,
and is then probably in a close binary.

As shown in Figure 3, all three variables show Ca II K and H
emission lines in the absorption core, indicating the presence of
active chromospheres. Ca II K and H emission lines are common among
cool dwarf and evolved stars, signaling the magnetic dynamo activity
in chromospheres (Kraft 1967; Soderblom 1983; Dupree \& Smith 1995).
Their fluxes correlate with stellar rotational velocities (see
Strassmeier et al. 1994; Pasquini et al. 2000 and references
therein). Followed Linsky et al. (1979), assuming $V-R=0.84$ for
K2III stars (Cox 2000), we calculated the net chromospheric loss in
the K lines, which are $\log \mathcal{F'}(K)=6.35$, 6.13 and 5.98,
indicating rotation velocities about 40 km/s, 25 km/s and 16 km/s
(Strassmeier et al. 1994; the uncertainty is large though, about 0.5
dex), for J0830, J0754 and J0736, respectively.

\begin{figure}
\epsfig{file=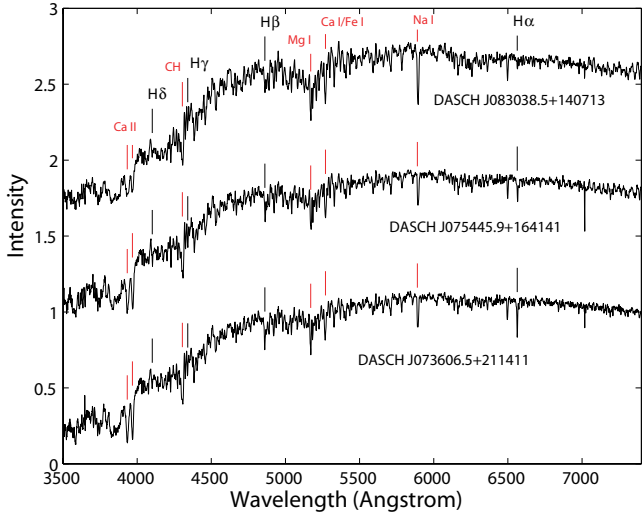, angle=0, width=\linewidth}
\caption{FAST 300 grating spectra of the 3 erratic variables with
lightcurves shown in Figure 1. All of them are K2III stars. Spectral
resolution is about 7\AA.}
\end{figure}

\begin{figure*}
\epsfig{file=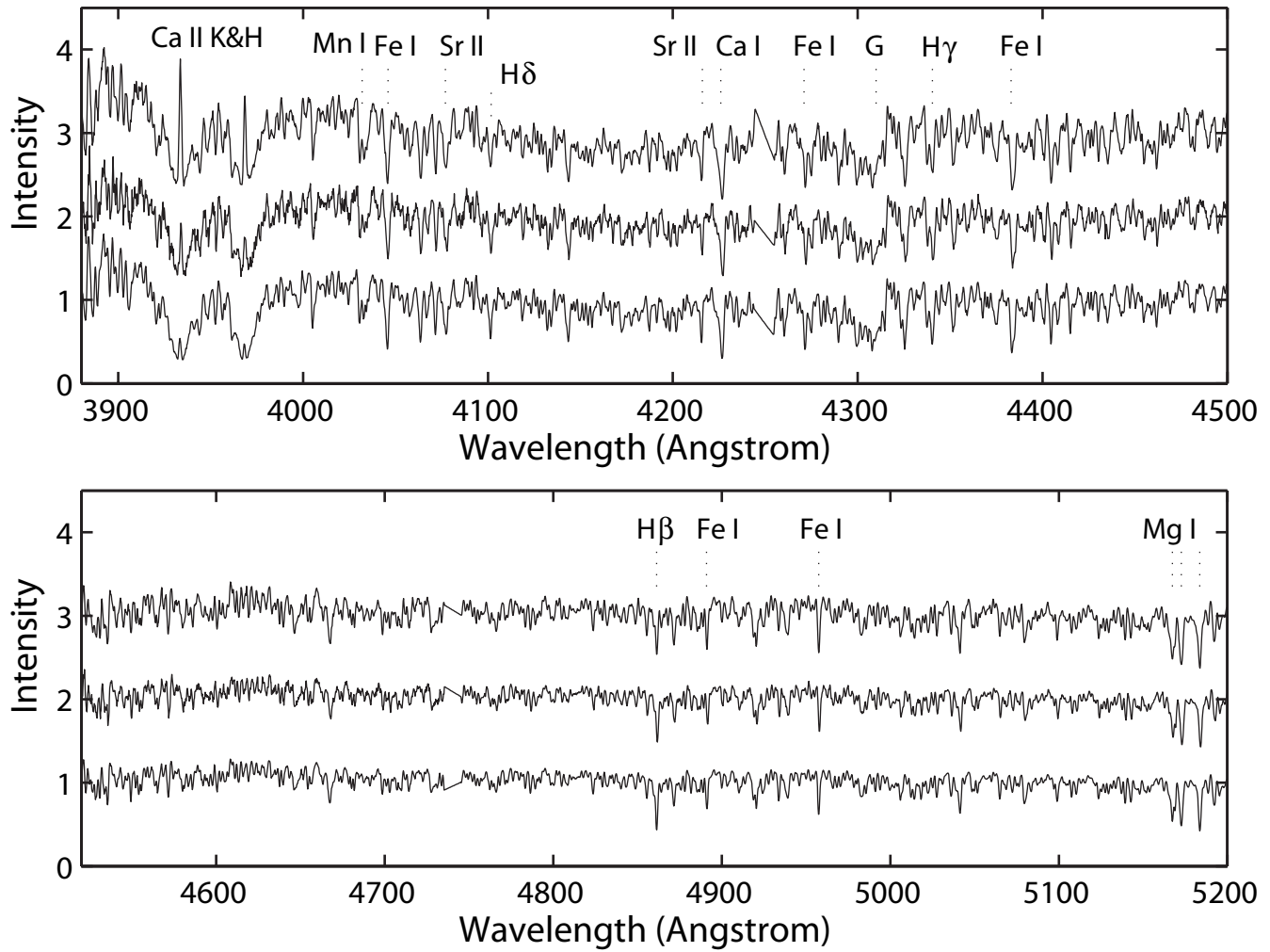, angle=0, width=\linewidth}
\caption{Gemini B1200 grating spectra of the 3 erratic variables.
From top to bottom are spectra of J0830, J0754, and J0736,
respectively. Continuum are removed by a sixth-order polynomial fit.
Spectral resolution is about 1.2 \AA\ ($R=3744$). }
\end{figure*}

\subsection{Occurrence Fraction of K2III Long-term Erratic Variables}
The plates we analyzed are centered around the open cluster M44. The
coverage decreases when the region is farther away from the center
of M44. There is roughly a $40\times40$ deg$^2$ region with good
coverage (i.e. more than 100 scans). From Keplercam photometry of a
3.2 deg$^2$ region near M44, we estimate there are about 4000-6000
stars with $12<B<13.5$ and $0.85<g-r<1.02$ in this $40\times40$
deg$^2$ region. If we assume that objects with $0.85<g-r<1.02$ are
K2III stars (Covey et al. 2007; Smith et al. 2002; note that some
dwarfs and supergiants are also included, but there are many more
giants than supergiants, and at 12-13 magnitude, we are seeing more
giants than dwarfs), then the event rate of such long-term variables
among K2III stars is about $3/5000\sim 0.06\%$.

\section{Discussion on Individual Objects}
Table 1 lists the galactic coordinates, GSC2.3.2 B-R colors, GSC B,
ASAS V and 2MASS JHK magnitudes, [Fe/H], proper motions, radial
velocities, distances, and galactic velocities for the three
variables. J0830 and J0754 have high proper motions ($>10$ mas/yr),
while J0736 has no proper motion within its error of $2$ mas/yr. The
3 stars are about $500-800$ pc above the galactic plane. J0830 and
J0754 have velocities about 100 km/s, while J0736 probably has
smaller velocity $<30$ km/s. Our spectra indicate metallicities
[Fe/H]$\sim -1$ to 0. They are most probably thick disk giants
(Carollo et al. 2009). The 2MASS $H-K$ color of J0830 is 0.27, which
is redder than $H-K=0.04$ expected for a K2 giant (Covey et al.
2007). Note that J0830 also showed large reddening in the 1970s and
1980s in our plates during its dimming, as shown in Figure 1. As
K2III stars with [Fe/H]$\sim -1$ to 0, the 3 erratic variables could
be on the Red Giant Branch (RGB), but not near the tip, on the
Horizontal Branch (HB), or early asymptotic giant branch (AGB).

Timescales of 10-100 year are too long to be pulsations driven by
ionization (Percy 2007), or convective instabilities which have
timescales in the order of a year for a K giant (Prialnik 2000). The
lightcurves of these 3 erratic variables look somewhat like R
Coronae Borealis (RCB) stars. RCB stars are rare, hydrogen
deficient, carbon-rich supergiants which undergo large amplitude
($3-8$ mag) fading events lasting weeks to a few years as dust
condensates to block the light along the line of sight (Clayton
1996). Our variables share some of these typical RCB lightcurve
features, and the lightcurve of J0830 is similar to some hot ($T\sim
20000$ K) RCB stars (De Marco et al. 2002). Dust obscuration events
have also been observed in some other carbon rich stars. Whitelock
et al. (2006) found about one-third of the Carbon-rich Miras and
other AGB stars undergo dimming episodes.

However, all of the three variables presented here are very
different from normal RCB stars in the following aspects, and
therefore are not RCB stars (or any other type of carbon-rich AGB
stars): They are not post-AGB supergiants like RCB stars (Alcock et
al. 2001); much cooler than most RCBs; have much longer variability
timescales and much smaller amplitudes than average RCB stars; have
no strong carbon absorption bands, and have strong hydrogen
absorption lines. J0830 has weaker H$\alpha$ absorption which might
due to the chromospheric emission in the core, since the Ca II
emission lines are very strong in J0830 and H$\alpha$ emission is
usually strongly correlated with Ca II lines (Cincunegui et al.
2007).

Some symbiotic stars also show irregular long-term variations (see
e.g. Sokoloski et al. 2006; Skopal 2008). However, there is no
symbiotic signature in our variables, such as nebula emission lines
or combination spectra. There is no common model that works well for
these 3 erratic variables, as we discuss below.

\subsection{J0830}{\footnote{Some of the models in this subsection may also apply to J0754 and
J0736}}
 The first possible model is dimming caused by obscuration of
dust shells ejected from the star, which is consistent with our
observation that the object became redder when it was fainter and
its redder 2MASS $H-K$ color. However, how to produce and maintain
the dust for such a long time is not clear. For RGB stars,
significant mass loss occurs only at the very tip (Origlia et al.
2002), which is much too luminous for the three erratic variables.
Bedding et al. (2002) found a 5-yr dimming event in the oxygen-rich,
Mira-like variable L$_2$ Pup, likely due to absorption by a dust
shell containing silicates. L$_2$ Pup is located near the tip of the
AGB, and thus has high mass-loss rate ($5\times10^{-7}\ M_{\odot}$
yr $^{-1}$), which if applied to our K2III stars would require they
have low metallicity and be halo stars. Tsuji (2009) found excess
absorption in CO lines of Arcturus (K1.5III), and proposed it might
be caused by the formation of molecular clouds in the outer
atmosphere. Arcturus only shows low amplitude optical variations of
a few percent (Bedding 2000) and thus the absorption due to such
molecular clouds is not enough to account for the 1 magnitude
dimming events in the three erratic variables.

A second possible model, is the lithium flash near the RGB
luminosity bump (Palacios et al. 2001), which proposed that rotation
induced mixing leads to a thin and unstable lithium burning shell,
which leads to an increase of nuclear luminosity, and mass loss
which might account for the formation of a dust shell around the
star and thus the reddening of the star. All of our 3 variables are
chromospherically active and probably fast rotators, which fit to
this model. However, the variation timescale and amplitude of
surface luminosity in this model is unknown.

A third possible model, which may be the most plausible one, is the
evolution phase when the star is leaving the HB and beginning to
ascend the AGB, at the point that helium is exhausted in the core
and helium burning is ignited in a surrounding shell. This could
apply to the point L in Figures 2 and 4 of Sackmann et al. (1993).
We plotted the evolutionary tracks for $[Z=0.008, Y=0.25]$
([Fe/H]$\sim$-0.4 if assume the same composition as the Sun) from
Girardi et al. (2000), and found stars with mass $0.8-1.2$ $M_\odot$
are crossing this evolutionary phase near $T\sim4400$ K and
$L\sim10^{2.2} L_{\odot}$ (which is the temperature and luminosity
for K2III stars; Cox 2000). The ignition of the helium-burning shell
causes expansion and makes the hydrogen-burning shell expand and
cool, which makes the surface luminosity decline and the color
redden, and then recover later. If this process is similar to
thermal pulses in He-burning shell of later AGB stars, then first
the surface luminosity decreases by a factor of 2 over 50-100 yr,
and then increases by a factor of 2 over 200 yr (Mowlavi 1999),
which is roughly consistent with J0830 and J0736. More theoretical
study and simulations of this evolution phase may clarify the nature
of our variables.

Two other interesting observations might be related to the three
erratic variables. Edmonds \& Gilliland (1996; hereafter EG96) found
a new class of variable stars in K giants with amplitudes $5-15$
mmag and periods of days clumped in the color-magnitude diagram with
B-V=1.1-1.2, which is the location of K2III stars. They lie on both
the AGB and the RGB. Bedin et al. (2000) found a `heap' in the
luminosity function which is about 1.4 magnitude brighter than the
RGB bump, and in the case of 47 Tuc, the heap is in a similar
location on the RGB to the variables found by EG96. The heap and the
new class variables are not well understood yet (see e.g. Salaris et
al. 2002), and it is surprising that they have similar color and
luminosity class as our erratic variables, which show much longer
timescale and larger amplitude variations. There might be some
intrinsic link between the objects in the three independent
observations. In Figure 4 of Sackmann et al. (1993), the location of
the He-shell ignition is just about 0.6 dex, i.e. 1.5 magnitude,
brighter than the RGB bump, consistent with the luminosity of K
giant variables in EG96.

\subsection{J0754}
This star is not yet included in any known variable catalogs. The
sharper decline and slower recovery are similar to what is seen in
RCB stars. It might be puffing off outer layers at irregular
intervals, which blocks the light from the star.

\subsection{J0736}
This variable is probably a binary. Given its possible orbital
period of $\sim\le15$ days ($\sim30$ km/s change in radial
velocities from Feb. 4 th to Feb. 19th) and its Ca K and H line
emission, it might share some properties in common with RS CVn
systems. It is not yet included in any known variable catalogs.
Similar to J0830 and J0736, it might be puffing off out layers which
blocks the light from the star, as supported by the measured
reddening during dimming; it might also be lithium flash near the
RGB bump, or the He-shell ignition in the lower AGB, as discussed in
section 3.1.

\begin{table*}[tb]
\caption{List of the three unusual long-term variables found in
DASCH. \label{tbl-1}} \centering \leavevmode
\begin{minipage}{\textwidth}
\tabcolsep 3.2pt \tiny
\begin{tabular}{lcccccccccccccccc}
\tableline\tableline DASCH Name  & Gal. & coord. &
GSC\tablenotemark{a} & GSC\tablenotemark{a} & ASAS\tablenotemark{b}
& & 2MASS\tablenotemark{c} & &[Fe/H] & PMRA\tablenotemark{d} &
PMDec\tablenotemark{d} & v$_{r}$\tablenotemark{e}  &
D\tablenotemark{f} & u\tablenotemark{g} &
v\tablenotemark{g} & w\tablenotemark{g}  \\
 & \emph{l} & \emph{b} & B & B-R & V & J &H & K & & mas/yr & mas/yr & km/s & kpc &  &  &  \\
\tableline J083038.5+140713  & 211 & 28 & 12.90 &
1.40 & 10.85 &8.82&8.20& 7.93& $-0.3\pm0.3$ & 9.0 & -10.8 & $0\pm8$ & 1.8 & -82 & -79 & 37  \\
J075445.9+164141   & 205 & 21 & 13.30& 1.61 & 12.24 &10.42&9.83&
9.70& $-0.9\pm0.4$ &-3.0&
-10.2   & $22\pm8$ & 2.1 & -2 & -81  & -48   \\
J073606.5+211411  & 198 & 19 & 12.66 & 1.29 & 11.19 &9.30&8.78&8.67
& $-0.9\pm0.4$ & -1.8
& -2.3 & -\tablenotemark{e} & 1.6 & - & - & -  \\
\tableline
\end{tabular}
\label{outcomes} \tablenotetext{a}{From GSC2.3.2 catalog (Lasker et
al. 1990).} \tablenotetext{b}{Median ASAS V magnitudes (Pojmanski
2002).} \tablenotetext{c}{2MASS J, H and K magnitudes with typical
uncertainty about 0.02 mag (Skrutskie et al. 2006).}
\tablenotetext{d}{Proper motion in RA and Dec, from Tycho-2 catalog
with typical uncertainty about 2 mas/yr (Hog et al. 2000).}
\tablenotetext{e}{Radial velocity. The radial velocity of J0736 is
variable.}
 \tablenotetext{f}{Distances of the objects assuming absolute B magnitude of 1.66 for K2III stars (Cox 2000).}
 \tablenotetext{g}{Galactic space velocity in km/s, corrected to the local standard of rest.
 u - positive toward the Galactic anti-center;
 v - positive in the direction of Galactic rotation;
 w - positive toward the North Galactic Pole. We didn't estimate the Galactic space velocity for J0736,
 due to the uncertain nature of its radial velocity and its small proper motion velocity which is consistent with zero within uncertainty.}
\end{minipage}
\end{table*}

\section{Summary}
We have found three very interesting long-term variables, which do
not resemble known classes of variables previously reported. We
found no model reported for red giants which could explain both
their timescales and amplitudes. The underlying causes of their
10-100 yr variations might be related to evolutionary nuclear
shell-burning instability and/or variable dust obscuration. Higher
resolution spectra and Infra-Red observation in future are needed to
constrain surface gravity, mass, and possible dust properties. More
theoretical work on the lithium flash near the RGB bump and the
helium shell ignition in the AGB, including surface luminosity
variation and possible dust formation, will be helpful to understand
their nature. As most likely thick disk stars with [Fe/H]$\sim -0.9$ to $-0.3$,
they are likely to have ages $8-16$ Gyr (Bensby et al. 2004),
and therefore likely to be $\sim0.8-1.1$ $M_\odot$ (Girardi et al. 2002).
If all 3 stars are $\sim1$ M$_\odot$ for which the RGB
is nearly vertical, their similar spectral type is expected. The
evolutionary timescales of variations in models 2 and 3 (section
3.1) are both in the range $\sim200-1000$ yr. A $\sim1$ M$_\odot$
star with Z=0.004 to 0.008 ([Fe/H]$\sim -0.7$ to $-0.4$)
spends $\sim10^{7.2}$ yr during RGB and $\sim10^{6.4}$ yr during AGB phases as K2 giants
(T$\sim$4300 to 4500 K, Cox 2000; Girardi et al. 2002). Therefore, the predicted rate of such
erratic variables among K2 giants is $\sim 600/10^{7.2}\sim0.004\%$. This is
$\sim15$ times smaller than our estimated occurrence fraction of our
K2III variables ($0.06\%$), indicating other mechanisms beyond the
evolutionary models we discussed in the paper might be relevant.

We did a general search for long-term variables and it turned out
surprisingly that three most interesting objects with drops in
timescales $\sim10-100$ year are all K2 giants. Is this a
coincidence or are thick disk K2 giants special? Due to the small
size of the sample, we cannot answer the question for sure yet. We
are working on a larger sample of similar long-term variables over
more plates, and hopefully we will have more knowledge about the
demographic very soon. We note that $\sim 1200$ plates covering M44
field are only $\sim 0.2\%$ of the whole Harvard plate collection,
and there is huge potential to find more interesting objects and new
classes of variables.

Stellar evolution proceeds on astronomical, not human, timescales
(except for pulsations and eruptive events). However,
evolution-driven changes on $\sim100$ yr timescales, such as shell
burning flashes and core helium flash in giants, are rare but can be
observed with the DASCH database with both long-term data coverage
and very large stellar samples. Note that AAVSO also provides
invaluable long-term data, but mostly for a few thousand much
brighter sources, see http://www.aavso.org. Instead of waiting for
another century to gather data to study 100-yr variability, we could
make it available for bright objects (B$<$15 mag) in several years
from DASCH, provided support for this full digitization scanning can
be found.

\acknowledgments We are grateful to the anonymous referee for
helpful comments and suggestions which significantly improved the
manuscript. We thank Alison Doane, Jaime Pepper, Bob Simcoe, George
Champine and Doug Mink at CfA for their work on DASCH. We thank Lars
Bildsten, Andrea Dupree, Geoffrey Clayton, Anna Frebel, Warren
Brown, Perry Belind and Peter Edmonds for enlightening and helpful
discussion. S.T. thanks FLWO staff for hospitality during the course
of this work, Mark Everret for reducing the Keplercam data, Nathalie
Martimbeau, Susan Tokarz and Bill Wyatt for the help on the
wavelength calibrations of FAST spectra. This research has made use
of the WEBDA, GSC 2.3.2 and 2MASS catalogs, SIMBAD, HEASARC and the
AAVSO database. The DASCH project gratefully acknowledges support
from NSF grants AST-0407380 and AST-0909073.

\end{document}